\documentclass[aps,prl,twocolumn,groupedaddress,preprintnumbers,nofootinbib]{revtex4-1}
\usepackage[dvips]{graphicx}
\usepackage[dvipsnames]{xcolor}
\usepackage{amsmath,amssymb,slashed,hyperref,mathtools,array,braket,cancel}
\usepackage[none]{hyphenat} 
\usepackage{float}
\usepackage{bbold}
\usepackage{comment} 
\usepackage{hyperref}
\hypersetup{
	colorlinks = true,
	linkcolor = Mahogany,
	anchorcolor = Mahogany,
	citecolor = Mahogany,
	filecolor = Mahogany,
	urlcolor = Mahogany
}
\newcommand{\nn}{\nonumber}
\newcommand{\sd}{\mathrm{d}}

\newcommand{\bb}[1]{\mathbb{#1}}
\newcommand{\cl}[1]{\mathcal{#1}}

\renewcommand{\to}{\longrightarrow}

\def\prd{\ref@{Phys.~Rev.~D}}        

\newcommand{\td}[1]{
	\if\notesOn1
	\todo[inline]{#1}
	\fi
}
\DeclareSymbolFont{matha}{OML}{txmi}{m}{it}
\DeclareMathSymbol{\varv}{\mathord}{matha}{118}
\def\notesOn{1}
\begin{document}
\title{Rotating Black Holes in Cubic Gravity}

\author{Daniel J. Burger$^\alpha$, William T. Emond$^\lambda$ and Nathan Moynihan$^\alpha$}
\email{nathantmoynihan@gmail.com}
\affiliation{$^\alpha$High Energy Physics, Cosmology \& Astrophysics Theory group\\
	and	The Laboratory for Quantum Gravity \& Strings\\\\
	Department of Mathematics and Applied Mathematics, University of Cape Town, Rondebosch, Cape Town 7700, South Africa\medskip\\
	$^\lambda$School of Physics and Astronomy, University of Nottingham,
	University Park, Nottingham NG7 2RD, United Kingdom}

\begin{abstract}
	Using on-shell amplitude methods, we derive a rotating black hole solution in a generic theory of Einstein gravity with additional terms cubic in the Riemann tensor. We give an explicit expression for the metric in Einsteinian Cubic Gravity (ECG) and low energy effective string theory, which correctly reproduces the previously discovered solutions in the zero angular-momentum limit. We show that at first order in the coupling, the classical potential can be written to all orders in spin as a differential operator acting on the non-rotating potential, and we comment on the relation to the Janis-Newman algorithm. Furthermore, we derive the classical impulse and scattering angle for such a black hole and comment on the phenomenological interest of such quantities.
\end{abstract}

\maketitle
\section{Introduction} We have recently entered the era of gravitational wave astronomy, where one of the key subjects of observation will be black holes \cite{Abbott:2016blz,TheLIGOScientific:2017qsa}. Analyses of the initial black hole data suggests further confirmation of General Relativity (GR) as the correct low-energy description of gravity. Having said this, there still remain pressing open issues in cosmology, such as the dark sector of the universe and more recently, the $H_0$--problem. 
From a theoretical viewpoint, it is reasonable to expect that GR is not the final say on gravity, and therefore it is prudent to study gravitational theories that extend beyond GR. This can be considered either in an effective field theory (EFT) sense, in which one considers the higher-order curvature terms expected from e.g. low-energy string theory, or simply as modified theories of gravity. In this paper, we shall study the effects of adding cubic curvature contributions to the Einstein-Hilbert action, specifically considering rotating black hole solutions to the field equations. Such analyses are important in order to test the validity of GR in the strong gravity regime of black holes, and further, to identify if such modifications to GR are necessary.

Since angular momentum is conserved, we would expect that nearly all astrophysical black holes in the universe will be spinning, irrespective of the specific theory of gravity under consideration. In a recent paper, two of the present authors used modern amplitude methods to derive non-rotating black hole solutions in cubic theories of gravity \cite{Emond:2019crr}. In this work, we extend this analysis to include rotating solutions by considering scattering amplitudes involving particles with arbitrary spin. 
In general, finding solutions to the field equations in covariant theories of gravity is fraught with difficulty, due to the non-linearity of gravity, the choice of coordinate systems and the sheer number of terms one often has to consider, not to mention the unwieldy complications of rotating solutions. Modern amplitude techniques have been extremely useful in this regard \cite{Holstein:2008sx,Neill:2013wsa,Vaidya:2014kza,Damour:2016gwp,Cachazo:2017jef,Cheung:2018wkq,Carballo-Rubio:2018bmu,Guevara:2018wpp,Chung:2018kqs,Caron-Huot:2018ape,Brandhuber:2019qpg,Cristofoli:2019ewu,Chung:2019duq,Guevara:2019fsj,Bern:2019crd,Arkani-Hamed:2019ymq,Moynihan:2019bor,Siemonsen:2019dsu,Kalin:2019rwq}, partly due to their manifest gauge invariance and partly to the technology that has been developed over the last decade to greatly simplify computations. One particularly efficient method to compute the classical (perturbative) solutions that arise from amplitudes is the so-called \textit{Leading Singularity} \cite{Cachazo:2017jef}, the highest codimension singularity of a given amplitude that allows one to efficiently compute the classical contributions by taking appropriate limits.

In theories of gravity, spin-effects can be found in post-Newtonian or post-Minkowskian multipole expansions \cite{Porto:2005ac,Levi:2010zu,Levi:2014gsa,Levi:2015msa,Damour:2016gwp,Vines:2016qwa,Vines:2017hyw,Vines:2018gqi,Levi:2018nxp}. At first post-Minkowskian order (1PM), classical effective matching gives rise to an all order in spin expansion \cite{Vines:2017hyw} in the two-body problem in general relativity. A similar analysis can be obtained directly from the amplitudes, as we will do here, by matching the results obtained from the amplitude calculations with an effective action \cite{Guevara:2018wpp,Guevara:2019fsj,Chung:2018kqs,Chung:2019duq}.

\begin{figure}[H]
	\centering
	\includegraphics[width=0.6\linewidth]{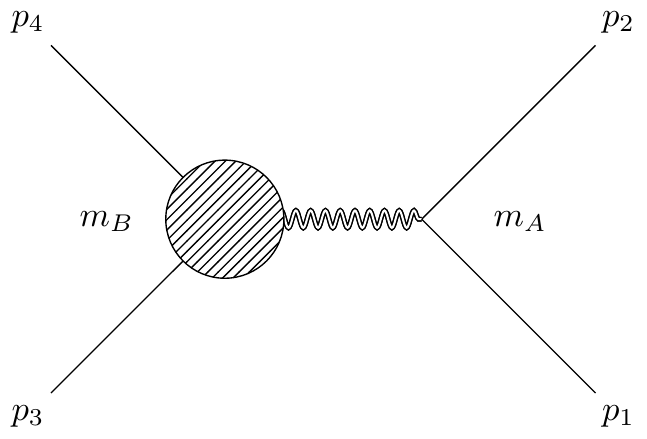}
	\caption{Gravitational probe of spinning particles in cubic gravity}
	\label{triangleLS1}
\end{figure}

We will consider a light scalar particle of mass $m_A$ and momentum $p_1$ probing the spacetime generated by a heavy spinning particle of mass $m_B$, momentum $p_3$ and spin $s$ (cf. the diagram given in Fig. \eqref{triangleLS1}).
In order to compute the classical contributions from such an interaction, we will compute the Leading singularities in the \textit{Holomorphic classical limit} (HCL) \cite{Guevara:2017csg}.
Cubic gravity is well studied in the literature \cite{Sisman:2011gz,Bueno:2016lrh,Bueno:2016xff,Hennigar:2016gkm,Hennigar:2017ego,Poshteh:2018wqy,Mir:2019rik}, and in this paper we will consider a generic six-derivative theory in four dimensions described by the action
\begin{equation}\label{key}
S = \int \sd^4x\sqrt{-g} \left(\frac{2}{\kappa^2}R + \lambda \mathcal{P}\right),
\end{equation}
where the coupling has mass dimension $[\lambda] = -2$ and
\begin{align}\label{eq:P def}
\mathcal{P}&=\beta_1 R^{\mu}_{\ \ \alpha\nu\beta}R^{\alpha \lambda\beta\sigma}R_{\lambda\mu\sigma}^{\ \ \ \ \nu}+\beta_2 R_{\mu\nu}^{\ \ \ \alpha\beta}R_{\alpha\beta}^{\ \ \ \lambda\sigma}R_{\lambda\sigma}^{\ \ \ \mu\nu}\nn\\ &+ \beta_3 R_{\mu\nu\alpha\beta}R^{\mu\alpha}R^{\nu\beta} + \beta_4 R_{\mu}^{\ \ \nu}R_{\nu}^{\ \ \alpha}R_{\alpha}^{\ \ \mu},
\end{align}
where the $\beta_i$ are generic coefficients which we will leave undetermined for the time being. In the particular case where $\beta_1=-\beta_3=12$ and $\beta_2=1$, $\beta_4=8$, the theory reduces to Einsteinian Cubic Gravity (ECG) \cite{Bueno:2016xff}. This recently constructed theory is of interest as an extension to GR that preserves the number of physical degrees of freedom in gravity. In fact, it is the unique extension of GR up to cubic order in curvature that propagates only massless spin-2 degrees of freedom on-shell (i.e. its linearized spectrum coincides with that of GR on maximally symmetric backgrounds) \cite{Bueno:2016xff}. 

Before proceeding, we should note that whilst we have kept the generic form of $\mathcal{P}$ for generality, it has been shown that the terms involving the Ricci curvature tensor do not contribute to on-shell amplitudes \cite{Metsaev:1986yb}.\footnote{This is due to the reparametrisation invariance of the S-matrix -- we can perform a field redefinition of the metric that removes all possible cubic curvature invariants other than those proportional to $\beta_3$ and $\beta_4$ in eq.\eqref{eq:P def} without affecting the on-shell amplitudes. Note that, in principle there are two further cubic terms that are a priori present in the action, however, due to the reason above, these also play no role on-shell. Since we are interested in ``ECG-like" theories of gravity, we neglect these additional terms in eq.\eqref{eq:P def}.} Therefore, the terms proportional to $\beta_3$ and $\beta_4$ in $\mathcal{P}$ will play no role in the following discussion. 
\subsection{HCL Parametrisation}
We will make heavy use of the HCL parametrisation, which we will review here along with our conventions and notation. The incoming and outgoing particles have momentum
\begin{equation}
p_1^2 = p_2^2 = m_A^2 \;,\quad p_3^2 = p_4^2 = m_B^2 \;,
\end{equation}
and the usual Mandelstam relations are given by
\begin{equation}
s \coloneqq (p_1 + p_3)^2 \;,\quad t \coloneqq (p_1 - p_2)^2 \;,\quad u \coloneqq (p_1 - p_4)^2 \;.
\end{equation}
Since we are computing scattering amplitudes on-shell, we will use spinor-helicity variables throughout, which make the little group properties of the amplitudes manifest. For massless particle momentum, we define the momentum bispinor as
\begin{equation}
    p^\mu_i\sigma_{\mu}^{\alpha\dot{\alpha}} = \lambda_i^\alpha\tilde{\lambda}^{\dot{\alpha}}_i~.
\end{equation}
The two-spinors $\lambda$ and $\tilde{\lambda}$ represent the positive and negative helicity states respectively, and contractions are performed using $\epsilon_{\alpha\beta}$ or $\epsilon_{\dot{\alpha}\dot{\beta}}$. Importantly, we will use angle and square bracket notation where contractions are denoted by
\begin{equation}
    \epsilon_{\alpha\beta}\lambda_i^\alpha\lambda_j^\beta \coloneqq \braket{ij},~~~~~\epsilon_{\dot{\alpha}\dot{\beta}}\tilde{\lambda}^{\dot{\alpha}}_i \tilde{\lambda}^{\dot{\beta}}_j \coloneqq [ij].
\end{equation}
For massless particles, the little group is $U(1)$, and the helicity of a given on-shell particle is given by simply scaling a given two-spinor by phase, where the amplitude satisfies
\begin{equation}
    \cl{M}(t\lambda_i,t^{-1}\tilde{\lambda}_i) = t^{-2h_i}\cl{M}(\lambda_i,\tilde{\lambda}_i)
\end{equation}
For massive particles, the little group is $SU(2)$ and so we need an additional index that transforms appropriately. We will use the formalism developed in \cite{Arkani-Hamed:2017jhn}, where massive particle momentum can be decomposed as
\begin{equation}
    p^\mu_i\sigma_{\mu}^{\alpha\dot{\alpha}} = \lambda_i^{I\alpha}\tilde{\lambda}^{J\dot{\alpha}}_i\epsilon_{IJ}~.
\end{equation}
We will, however, adopt the bold notation used in \cite{Arkani-Hamed:2017jhn}, supressing the massive little group indicies $I,J,K...$ in favour of bold angle and square bracket notation, e.g.
\begin{equation}
    \epsilon_{\alpha\beta}\lambda_i^{I\alpha}\lambda_j^{J\beta} \coloneqq \braket{\textbf{ij}},~~~~~\epsilon_{\dot{\alpha}\dot{\beta}}\tilde{\lambda}^{I\dot{\alpha}}_i \tilde{\lambda}^{J\dot{\beta}}_j \coloneqq [\textbf{ij}],
\end{equation}
where it is understood that the bold notation implies that each bold spinor has an index which can be restored from the fact that all indices must be fully symmetrised over.

Having dispensed with the relevant notation, we can define the massless exchanged momentum as
\begin{equation}\label{key}
K \coloneqq p_1 - p_2 = |\lambda]\bra{\lambda} = (0,\textbf{q}),~~~~~K^2 = t = -|\textbf{q}|^2,
\end{equation}
and work in a parametrisation that makes the classical pieces explicit, e.g.
\begin{equation}
\begin{split}
p_{3} & =|\eta]\langle\lambda|+|\lambda]\langle\eta|\,,\\
p_{4} & =\beta|\eta]\langle\lambda|+\frac{1}{\beta}|\lambda]\langle\eta|+|\lambda]\langle\lambda|\,,\\
\frac{t}{m_{B}^{2}} & =\frac{(\beta-1)^{2}}{\beta}\,,\\
\langle\lambda\eta\rangle & =[\lambda\eta]=m_{B}\,.
\end{split}
\label{eq:param}
\end{equation}
Using these variables, the classical limit of an amplitude is given by taking the $\beta\longrightarrow 1$ limit.
We also define the useful parameters
\begin{equation}
u \coloneqq [\lambda|p_1|\eta\rangle \;,\quad v \coloneqq [\eta|p_1|\lambda\rangle \;,
\end{equation}
with $u+v = 2p_1\cdot p_3$ in the HCL.
These are related to the mandelstam variables in the HCL via
\begin{align}
u &= m_Am_B(\rho + \sqrt{\rho^2 - 1}),\nn\\
v &= m_Am_B(\rho - \sqrt{\rho^2 - 1}),
\end{align}
where $$m_Am_B\sqrt{\rho^2 - 1} = \sqrt{(s - (m_A + m_B)^2)(s - (m_A - m_B)^2)}$$ and the non-dispersive limit is given by $\rho\rightarrow 1$.

Given these definitions and kinematics, we find the following relations
\begin{align}
\bra{\lambda}p_1|\lambda] &= -\frac{(\beta-1)^2}{\beta}m_B^2 + (1-\beta)(v-\frac{u}{\beta}),\nn\\
\bra{\eta}p_1|\eta] &= \frac{uv - m_A^2m_B^2}{(u-v)(\beta - 1)} + \cl{O}(\beta-1)^0\;.
\end{align}
Spin effects are given in terms of the Pauli-Lubanski spin-vector, which we will use in its mass-rescaled form, given by
\begin{equation}\label{key}
a^\mu = -\frac{1}{m^2}(P^\nu_i \bar{\sigma}_{\mu\nu}).
\end{equation}
The spin dependence is characterised by identifying \cite{Holstein:2008sx}
\begin{equation}\label{spinid}
\epsilon_{\mu\nu\rho\sigma}p_1^\mu p_3^\nu K^\rho a^\sigma = (E_A + E_B)(\textbf{a}\cdot \textbf{p}\times \textbf{q}),
\end{equation}
where $\textbf{p}$ is the relative momentum. 
We will work in the anti-chiral HCL basis, meaning that we choose a basis that contains only dotted indices (we can always exchange dotted for undotted indices using the momentum bispinor). Working in such a basis allows us to efficiently extract the spin dependence of the amplitudes that is otherwise obscured in the mixed basis. For example, the spin-dependence of a three particle amplitude with two spinning particles coupled to a massless particle of (positive) helicity $h$ can be exposed by a change of basis of the form
\begin{align}\label{key}
\cl{M}_3^{s,h} &= g(mx)^h\frac{\braket{\textbf{12}}^{2s}}{m^{2s}}\nn\\ &= -g(mx)^h\left[[\textbf{1}|\left(1 - \frac{|3][3]}{mx}\right)|\textbf{2}]\right]^{2s},
\end{align}
where the second term on the right hand side represents the spin contribution and is the on-shell avatar of the Gordon decomposition \cite{Holstein:2008sx,Moynihan:2019bor}.

For convenience, we will also define the relative momentum-dependence of the amplitudes in terms of the rapidity $w$. By working in the anti-chiral basis, we lose any information that might have been contained in the polarisation tensors of the external spinning particles, which are only defined in the mixed representation. In order to restore this lost information, we will turn to the \textit{Generalised Expectation Value} (GEV) introduced in Ref. \cite{Guevara:2018wpp}, which in our case amounts to normalising the amplitude via
\begin{equation}
\braket{\cl{M}} = e^{-K\cdot a}\cl{M}.
\end{equation}
Armed with the HCL parametrisation and this normalisation, we can compute the amplitudes relevant to derive the classical potential.
\section{Tree Level LS}
\begin{figure}[H]
	\centering
	\includegraphics[]{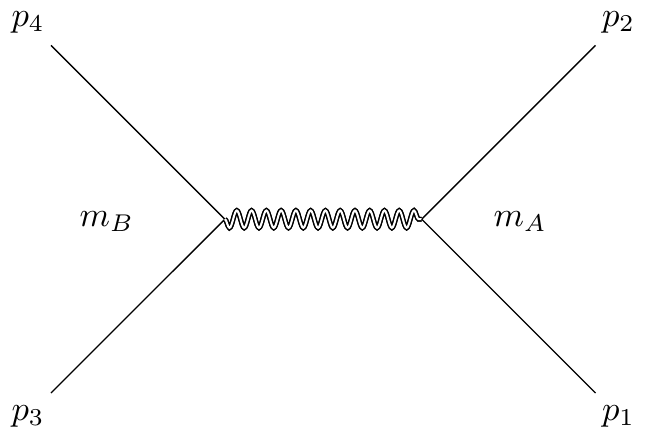}
	\caption{Tree level graviton exchange}
	\label{treediag}
\end{figure}

At tree level, there is no $\cl{O}(\lambda)$ contribution, and we need only to consider the diagram in Fig. \eqref{treediag}
The minimally coupled three-particle amplitude with one graviton and two spin-$s$ particles is given by \cite{Arkani-Hamed:2017jhn}
\begin{align}\label{key}
\cl{M}_3[1,2,K^{+2}] &= \frac{\kappa}{2}(mx_{12})^2\frac{\braket{\textbf{12}}^{2s}}{m^{2s}}\nn\\
\cl{M}_3[1,2,K^{-2}] &= \frac{\kappa}{2}\left(\frac{m}{x_{12}}\right)^2\frac{[\textbf{12}]^{2s}}{m^{2s}},
\end{align}
where $x_{ij}$ is defined via 
\begin{equation}\label{key}
x_{ij}\lambda_i^\alpha = \frac{\tilde{\lambda}_{i\dot{\alpha}}p^{\dot{\alpha}\alpha}_j}{m},~~~~~\frac{\tilde{\lambda}_i^{\dot{\alpha}}}{x_{ij}} = \frac{p_j^{\dot{\alpha}\alpha}\lambda_{i\alpha}}{m}.
\end{equation}
The tree-level leading singularity is simply the residue in the $t$ channel, in this case given by the product of the above amplitudes. Considering the HCL parametrisation and working in the chiral representation, this is given by 
\begin{align}\label{kerr4pt}
\cl{M}_4^s &= -\left(\frac{\kappa}{2}\right)^2\frac{m_A^2m_B^2}{t}\left(\frac{x_{34}^2}{x_{12}^2}\left(\bb{1} + \frac{K\cdot a}{s}\right)^{2s} + \frac{x_{12}^2}{x_{34}^2}\right),\nn\\
&= -\left(\frac{\kappa}{2}\right)^2\frac{1}{t}\left(u^2\left(\bb{1} + \frac{K\cdot a}{s}\right)^{2s} + v^2\right),
\end{align}
where we have formally defined the spin vector $a = 2s\tilde{a}$ and noted that
\begin{equation}\label{key}
u = m_Am_B\frac{x_{34}}{x_{12}},~~~~~v = m_Am_B\frac{x_{12}}{x_{34}}.
\end{equation}

For spin zero, in the non-relativistic limit, we find then the amplitude
\begin{equation}\label{key}
\cl{M}_4^0 = -\left(\frac{\kappa^2}{2}\right)\frac{m_A^2m_B^2}{t}~.
\end{equation}
\section{One Loop LS}
\begin{figure}[H]
	\centering
	\includegraphics[]{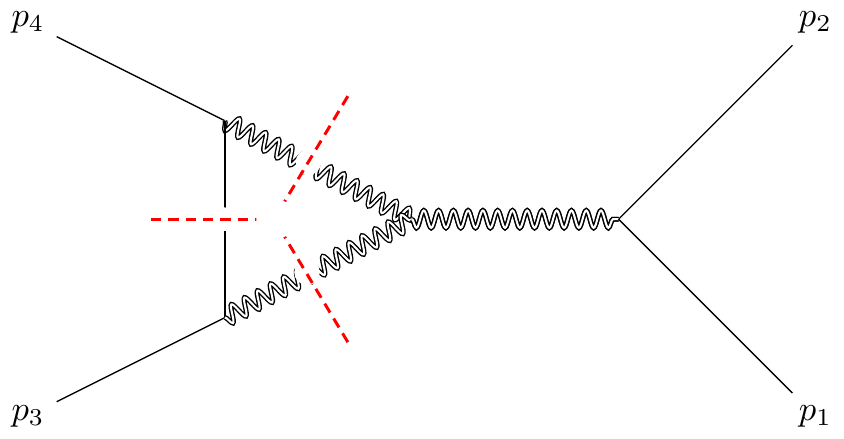}
	\caption{One-loop leading singularity}
	\label{triangleLS}
\end{figure}

In order to derive a black hole solution at order $\cl{O}(\lambda)$ for cubic gravity, we only need to consider one loop diagram given in Fig. \eqref{triangleLS}.
\vspace{-0.6em} 
We label the internal graviton momentum $k_3$ and $k_4$ and the massive internal scalar $L$. We cut every internal propagator and integrate over the internal momentum, which will now be fully localized since we are only considering the LS. 
\begin{equation}\label{key}
L = z\ell + \omega K,~~~~~|\ell] = |\eta] + B|\lambda],~~~~~\bra{\ell} = \bra{\eta} + A\bra{\lambda}.
\end{equation}
Demanding the on-shell cut conditions $k_{3,4}^2 = L^2 - m_B^2$ (and imposing $\ell^2=0$) fixes $\omega = -\frac{1}{z}$ with $A = -B = -\frac{1}{z}\frac{2\beta}{1+\beta}$. Given this parametrisation, the LS is given by a contour integral where the integrand is a product of tree-level on-shell amplitudes 
\begin{equation}\label{key}
\cl{I} = \frac{1}{16\sqrt{-t}m_B}\oint_{\Gamma}\frac{\sd y}{y}\cl{M}_3^+\cl{M}_3^+\cl{M}_4^{--},
\end{equation}
where we have taken the $\beta\longrightarrow 1$ limit.

We will again use the HCL parametrisation to express the internal momentum spinor helicity variables as
\begin{align}\label{key}
|k_3] &= \frac{1}{\beta+1}\left(|\eta](\beta^{2}-1)y+|\lambda](1+\beta y)\right),\nonumber\\
\bra{k_3} &= \frac{1}{\beta+1}\left(\langle\eta|(\beta^{2}-1)-\frac{1}{y}\langle\lambda|(1+\beta y)\right),\nonumber\\
|k_4] &= \frac{1}{\beta+1}\left(-\beta|\eta](\beta^{2}-1)y+|\lambda](1-\beta^{2}y)\right),\nonumber\\
\bra{k_4} &= \frac{1}{\beta+1}\left(\frac{1}{\beta}\langle\eta|(\beta^{2}-1)+\frac{1}{y}\langle\lambda|(1-y)\right).
\end{align}
\subsection{Tree-Level Components}
We now work out the tree-level components that go into the loop, expressed in terms of the HCL parameters. For a particle of generic spin $s$ emitting a graviton, the three-particle amplitude is given by
\begin{align}\label{key}
\cl{M}_3[1^s,2^s,K^{+2}] &= -\frac{\kappa}{2}m^2x^2_{12}\left(\bb{1} + \frac{K\cdot a}{s}\right)^{2s},\\
\cl{M}_3[1^s,2^s,K^{-2}] &= -\frac{\kappa}{2}\left(\frac{m^2}{x^2_{12}}\right),
\end{align}
where we have chosen to work in the purely anti-chiral basis and expose the spin dependence of positive helicity amplitudes. We can therefore work out the product of three-particle amplitudes that go into the LS 
\begin{align}\label{key}
&\cl{M}_3[3^s,-L^s,k_3^{+2}]\cl{M}_3[4^s,L^s,k_4^{+2}] = \left(\frac{\kappa}{2}\right)^2m_B^4 y^4\nn\\&\times\left(1 + \frac{(1+y)^2}{4y}\frac{K\cdot a}{s}\right)^{2s}\left(1 - \frac{(1-y)^2}{4y}\frac{K\cdot a}{s}\right)^{2s},
\end{align}
where we have used the fact that $x_{3L} = x_{4L} = -y$ in this parametrisation. 

We can also determine the four-point that goes into the LS in terms of these parameters. Noting that we can express
\begin{align}
     k_3.p_1 &= k_3.p_2 + \cl{O}(\beta - 1)^2= (\beta - 1)\frac{\left(1-y^2\right) (v-u)}{8 y},
\end{align}
we find
\begin{widetext}
\begin{align}
\cl{M}_4[p_1,p_2,k_3^{-2},k_4^{-2}] &=\frac{3}{16}\frac{\kappa^4\lambda\braket{k_3k_4}^4}{(p_1-p_2)^2}\Big[\beta_1\left((k_3\cdot p_1 + k_3\cdot p_2)^2 - m_A^2k_3\cdot k_4\right) + 8\beta_2(k_3\cdot p_1)(k_3\cdot p_2)\Big]
\nn\\[0.8em] 
&= -\frac{3}{64}\frac{\kappa^4\lambda}{t}\frac{m_B^4(\beta - 1)^6}{y^4}\Big[2\beta_1m_A^2m_B^2 - (\beta_1 + 2\beta_2)\frac{\left(1-y^2\right)^2 (v-u)^2}{4 y^2}\Big] \nn\\[0.8em] 
&= -\frac{3}{64}\kappa^4\lambda\frac{t^2}{m_B^2y^4}\Big[2\beta_1m_A^2m_B^2 - (\beta_1 + 2\beta_2)\frac{\left(1-y^2\right)^2 (v-u)^2}{4 y^2}\Big] \;.
\end{align}
Putting this all together, we find the LS to be evaluated is
\begin{align}
\cl{I}^s &= -\frac{3}{4096}\kappa^6\lambda m_B(-t)^{3/2}\oint\frac{\sd y}{y}\left(1 + \frac{(1+y)^2}{4y}\frac{K\cdot a}{s}\right)^{2s}\left(1 - \frac{(1-y)^2}{4y}\frac{K\cdot a}{s}\right)^{2s}\nn\\&\qquad\qquad\qquad\qquad\qquad\qquad\times\Big[2\beta_1m_A^2m_B^2 - (\beta_1 + 2\beta_2)(v - u)^2(1 + y)^2\Big] \;.\label{mainLS}
\end{align}
\end{widetext}
In order to check that this is the correct expression, we first choose $s = 0$ to ensure we are able to recover the results obtained in \cite{Emond:2019crr}. Taking the first term in eq. \eqref{mainLS}, and summing together with $m_A\leftrightarrow m_B$, we find
\begin{equation}\label{key}
\cl{I}_{\beta_1}^{s=0} = -\frac{3}{1024}\beta_1\kappa^6\lambda m_A^2m_B^2(m_A + m_B)\textbf{q}^3.
\end{equation}
This precisely matches the small $t$ expansion of the pure $\beta_1$ term of eq. (5.2) in \cite{Emond:2019crr}. Moving on to the second term, we find that the LS is
\begin{align}\label{key}
\cl{I}_{\beta_1 + 2\beta_2}^{s=0} &= -\frac{3}{4096}(\beta_1+2\beta_2)\kappa^6\lambda (m_A +  m_B)(v-u)^2\textbf{q}^3 \nn\\
&= -\frac{3}{1024}(\beta_1+2\beta_2)\kappa^6\lambda (m_A +  m_B)^3\textbf{p}^2\textbf{q}^3,
\end{align}
which agrees with the pure $(\beta_1+2\beta_2)$ term of eq. (5.2) in \cite{Emond:2019crr}. This leaves us confident that eq. \eqref{mainLS} is indeed the correct expression, and thus we shall move on to consider cases in which $s\neq 0$.
\section{All Order In Spin Classical Potential}
In momentum space, the classical potential is a related to the amplitude $\cl{M}$ by
\begin{equation}\label{key}
V(\textbf{q},\textbf{p}) = \frac{\braket{\cl{M}}}{4E_AE_B},
\end{equation}
where $E_i = \sqrt{m_i^2 + \textbf{p}^2 + \frac{\textbf{q}^2}{4}}$.

A consistent classical limit is only reached by taking $s\longrightarrow \infty$ as $\hbar\longrightarrow 0$ keeping $s\hbar$ fixed, since the intrinsic angular momentum of a spin $s$ particle is proportional to $s\hbar$ \cite{Maybee:2019jus}. We now make a further identification for the variables $u$ and $v$ as being
\begin{align}\label{key}
u = m_Am_Be^w,~~~~~v = m_Am_Be^{-w},
\end{align}
where $w$ is the rapidity.
Plugging this into the tree-level four point and taking the infinite spin limit gives
\begin{equation}\label{key}
\cl{M}_4^\infty = -\frac{\kappa^2}{4}\frac{m_A^2m_B^2}{t}(e^{2w}e^{2K\cdot a} + e^{-2w})~.
\end{equation}
After normalisation with the GEV, we then find
\begin{equation}\label{key}
\braket{\cl{M}_4^\infty} = -\frac{\kappa^2}{4}\frac{m_A^2m_B^2}{t}\left(e^{2w}e^{K\cdot a} + e^{-2w}e^{-K\cdot a}\right).
\end{equation}
Doing the same for the loop amplitude yields
\begin{equation}
\cl{I}^\infty = -\frac{3\kappa^6\lambda}{2048}(-t)^{3/2} m_A^2m_B^3\beta(w)(1 + e^{2K\cdot a}),
\end{equation}

where we have defined $\beta(w) \coloneqq \beta_1 + 2(\beta_1 + 2\beta_2)\sinh^2 w$.
Again normalising this using the GEV gives
\begin{align}
\braket{\cl{I}^\infty} &= -\frac{3\kappa^6\lambda}{1024} m_A^2m_B^3\beta(w)\textbf{q}^3\cosh \textbf{q}\cdot \textbf{a}.\label{mainLSinfspinlimit}
\end{align}
The momentum space potential is therefore given by
\begin{align}
V &= \frac{\kappa^2}{16}\frac{m_Am_B}{\textbf{q}^2}\left(e^{2w}e^{\textbf{q}\cdot \textbf{a}} + e^{-2w}e^{-\textbf{q}\cdot \textbf{a}}\right)\nn \\&~~~-\frac{3\kappa^6\lambda m_Am_B^2}{4096}\beta(w)\textbf{q}^3\cosh \textbf{q}\cdot \textbf{a}.
\end{align}
Since we want to compare with the non-dispersive Kerr black hole solution we at this point restrict the potential to the non-dispersive terms (i.e. $w = 0$) and taking the Fourier transform, we find then that the all-order in spin potential is given by
\begin{align}\label{key}
V(r) &= -\frac{\kappa^2}{32\pi}m_Am_B\cos \left(\textbf{a}\cdot \nabla\right)\left(\frac{1}{r} - \frac{9\kappa^4\lambda\beta_1}{32\pi} \frac{m_A+m_B}{r^6}\right),\nn\\
&= -Gm_Am_B\cos \left(\textbf{a}\cdot \nabla\right)\left(\frac{1}{r} -288\pi G^2\lambda\beta_1 \frac{m_A+m_B}{r^6}\right).
\end{align}
We note that this form of the potential allows us to interpret the attachment of spin-factors to the amplitudes generating the spacetime as the on-shell avatar of the Newman-Janis algorithm for higher derivative gravity, in precisely the same way as it does in the Kerr and Kerr-Newman cases \cite{Arkani-Hamed:2019ymq,Moynihan:2019bor}. This is because attachment of such factors gives rise to a differential operator which performs a complex deformation of the coordinates $r\rightarrow r + ia$ (a simple extension of the translation operator), i.e.
\begin{align}
\cos(a\cdot\nabla)f(r) &= 2\Re\left(f(r+ia)\right),\\
\sin(a\cdot\nabla)f(r) &= 2\Im\left(f(r+ia)\right),
\end{align}
for any holomorphic function $f(r)$.
\section{Metric Construction}
In momentum space, the classical potential for a gravitomagnetic system is of the form
\begin{equation}\label{key}
V(\textbf{q}) = m\Phi(\textbf{q}) + \textbf{p}\cdot\textbf{A},
\end{equation}
where $\Phi$ is the gravito-electric field, $\textbf{A}$ is gravito-magnetic vector potential and $\textbf{p}$ is kinetic momentum. The perturbative metric can be decomposed in terms of these fields as
\begin{equation}\label{metricdecomp}
h_{00} = 2\Phi,~~~~~h_{0i} = -A_i,~~~~~h_{ij} = 2\Phi\delta_{ij}.
\end{equation}
We can then identify the relevant components of the metric
\begin{equation}\label{key}
\Phi = \lim_{m_A\longrightarrow 0}\frac{1}{m_A}V,~~~~~~ A_i = \lim_{m_A\longrightarrow 0}2\frac{\partial V}{\partial p^i},
\end{equation}
where we note that we will often write $p_i = m_Au_i$ in the limit we are interested in.

In order to identify the spin components, we are required to identify the dispersive terms that multiply $K\cdot a$ via
\begin{equation}\label{key}
m_Am_B\sinh w ~K\cdot a = i\epsilon_{\mu\nu\rho\sigma}p_1^\mu p_3^\nu K^\rho a^\sigma.
\end{equation}
In the centre of mass frame, we can use this on-shell condition to write
\begin{equation}\label{key}
K\cdot a = \frac{i\epsilon_{\mu\nu\rho\sigma}p_1^\mu p_3^\nu K^\rho a^\sigma}{m_Am_B\sinh w} = -i\textbf{u}\cdot (\textbf{a}\times \textbf{q}),
\end{equation}
where $\textbf{u}$ is the relative four velocity.

With this in hand, we can therefore rewrite the potential, keeping only the necessary dispersive terms, as
\begin{align}
V &= \cos (\textbf{u}\cdot (\textbf{a}\times \textbf{q}))\left[\frac{\kappa^2}{8}\frac{m_Am_B}{\textbf{q}^2} -\frac{3\kappa^6\lambda m_Am_B^2}{4096}\beta(w)\textbf{q}^3\right].
\end{align}
We find then, in momentum space,
\begin{align}\label{key}
\Phi &= \cosh(\textbf{q}\cdot \textbf{a})\left(\frac{\kappa^2}{8}\frac{m_B}{\textbf{q}^2} -\frac{3}{4096}\beta(w)\kappa^6\lambda m_B^2\textbf{q}^3\right),\\
A_i &= -\sinh(\textbf{q}\cdot \textbf{a})\times\nn\\
&~~~~~\left[\frac{\kappa^2}{4}\frac{m_B}{\textbf{q}^2} -\frac{3\kappa^6\lambda m_B^2}{2048}\beta(w)\textbf{q}^3\right](i\textbf{a}\times \textbf{q})_i \;,
\end{align}
which in position space is
\begin{flalign}\label{eq:Phi A_i}
    \Phi &= \cos(\textbf{a}\cdot\nabla)\left(\frac{Gm_B}{r}-\frac{288G^3\pi\lambda\beta(w)m_B^2}{r^6}\right),\nn\\
    &= \cos(\textbf{a}\cdot\nabla)\left(\Phi_{Kerr} + \Phi_{R^3}\right),\\
    A_i &= -\sin(\textbf{a}\cdot\nabla)\left(\frac{2Gm_B}{r} - 3456 \frac{G^3\pi\lambda\beta(w)m_B^2}{r^6}\right)\frac{(\mathbf{a}\times\mathbf{r})_i}{r^2} \nn\\
    &= -2\sin(\textbf{a}\cdot\nabla)\left(\Phi_{Kerr} + 6\Phi_{R^3}\right)\frac{(\mathbf{a}\times\mathbf{r})_i}{r^2},
\end{flalign}
from which the components of the metric can be determined. Specialising now to the case of ECG \cite{Bueno:2016xff,Bueno:2016lrh}, we choose $\beta_1 = 12,~\beta_2 = 1$ and $\lambda = -\frac{G\tilde{\lambda}}{16\pi}$, such that we find the following perturbative expansion for the metric
\begin{flalign}\label{eq:metric comps}
g_{00}^{ECG} &= 1-\frac{2GM}{r} - 432\frac{G^4M^2\tilde{\lambda} }{r^6} + \cdots \\ g_{0i}^{ECG} &= \left(1 + \frac{2GM}{r} + 2592\frac{G^4 M^2\tilde{\lambda}}{r^6}\right)\frac{(\mathbf{a}\times\mathbf{r})_i}{r^2}  + \cdots\nn\\
g_{ij}^{ECG} &= \left(1+\frac{2GM}{r} + 432\frac{G^4M^2\tilde{\lambda} }{r^6}\right)\delta_{ij} + \cdots\nn \;.
\end{flalign}
Note that these results found for $\Phi$, $A_i$ \eqref{eq:Phi A_i}, and the metric components \eqref{eq:metric comps}, reduce to the previously known non-rotating counterparts in the zero angular momentum limit. Indeed, in the limit $\mathbf{a}\to 0$, we find that $\Phi=\Phi_{Kerr} + \Phi_{R^3}$ and $A_i=0$, and correspondingly $g_{0i}^{ECG}=0$, which is in exact agreement with the results found in \cite{Emond:2019crr,Bueno:2016lrh}. Furthermore, taking the $\lambda \rightarrow 0$ limit gives the all order in spin Kerr metric \cite{Vines:2017hyw}.
\section{Classical Impulse and Scattering Angle}
We can also derive the classical impulse imparted to scalar probe particle, given in terms of the scattering amplitude by \cite{Kosower:2018adc}
\begin{align}\label{key}
\Delta p^\mu_1 &= \frac{1}{4m_Am_B}\int \hat{d}^4\bar{q}\hat{\delta}(\bar{q}\cdot u_1)\hat{\delta}(\bar{q}\cdot u_3)e^{-i\bar{q}\cdot b}i\bar{q}^\mu \braket{\cl{M}^\infty}.
\end{align}
The impulse is given in terms the incoming probe particle momentum $p_1 = m_Au_1$ and its colliding partner $p_3 = m_Bu_3$, and is simply a measure of the total change in momentum of particle 1 as a result of the collision. The impact parameter $b^\mu$ is a spacelike vector orthogonal to $u_1$ and $u_3$, and the delta functions ensure that we pick out the plane in which this lives. Plugging in $\braket{\cl{M}^\infty} = \braket{\cl{M}^\infty_4} + \braket{\cl{I}^\infty}$ and performing the integral transforms, we find
\begin{align}\label{impulse}
\Delta p_1^\mu &= -\frac{2Gm_Am_B}{\sinh w}\Re\Bigg[\frac{\tilde{b}_\perp^\mu \cosh 2w + 2i\cosh w \zeta^\mu}{|\tilde{b}_\perp|^2}\\ &+ 135G^2\lambda\pi^2 m_B\left(\frac{\beta_1\tilde{b}_\perp^\mu + 2i(\beta_1 + 2\beta_2)\sinh w\zeta^\mu}{|\tilde{b}_\perp|^7}\right)\Bigg]\nn
\end{align}
where $\zeta^\mu \coloneqq \epsilon^{\mu\nu\rho\sigma}\tilde{b}_\nu u_{1\rho} u_{3\sigma}$ and $\tilde{b}_\perp = b_\perp+ i\Pi a$, in which $\Pi^\mu_{\,\nu}= \epsilon^{\mu\alpha\rho\alpha\beta}\epsilon_{\nu\rho\gamma\delta}\frac{u_{1\alpha}u_{3\beta}u_1^\gamma u_3^\delta}{\gamma^2-1}$ is the projector into impact parameter space (i.e. into the plane orthogonal to both incoming velocities $u_1$ and $u_3$) \cite{Maybee:2019jus}. 

We can also derive the scattering angle, given in terms of the LS by \cite{Bjerrum-Bohr:2018xdl,Guevara:2018wpp,Bjerrum-Bohr:2019kec}
\begin{equation}\label{angle}
\theta \simeq 2 \sin\left(\frac{\theta}{2}\right) = \frac{-E}{(2 m_A m_B \sinh w)^2} \frac{\partial}{\partial b} \braket{\cl{M}^\infty(b)},
\end{equation}
where $\braket{\cl{M}^\infty(b)}$ is the LS in impact parameter space.
The tree-level LS is given by
	\begin{equation}
\braket{\cl{M}^\infty_4(b)} = \left(\frac{\kappa^2}{8\pi}\right) m_A^2 m_B^2 \sum_{\pm} e^{\pm 2 w} \ln|b\pm a|, 
\end{equation}
and the loop LS \eqref{mainLSinfspinlimit} by
	\begin{equation}
	\braket{\cl{I}^\infty(b)}  = -\frac{27\kappa^6\lambda}{4096\pi} m_A^2m_B^3 \sum_{\pm}\frac{\beta(w)}{(b\pm a)^5}.
	\end{equation}
	Plugging these back into \eqref{angle} we find the angle
	\begin{align}
	\theta &= \frac{- G E}{\sinh^2w}\sum_{\pm}\Bigg(\frac{e^{\pm 2 w}}{b\pm a} + 270\pi^2G^2 \lambda m_B\frac{\beta(w)}{(b\pm a)^6}\Bigg).
	\end{align}
It can also be checked that this angle is reproduced directly from the classical impulse in eq. \eqref{impulse}, by considering \cite{Vines:2017hyw}
\begin{equation}\label{key}
\theta = \frac{\Delta p_\perp}{|p_\perp|},
\end{equation}
where $\Delta p^\mu = \Delta p_\perp \hat{b}^\mu$ and $|p_\perp| = m\sinh w$.
\section{Rotating stringy black holes}
Having so far considered a generic cubic theory of gravity given by the action eq.~\eqref{key}, we can naturally extend our analysis to consider the corrections to the potential that arise from the $\alpha'^2$ part of low-energy four-dimensional effective action in bosonic string theory (with a constant dilaton), given by \cite{Metsaev:1986yb} 
\begin{equation}
    S = -\frac{2\alpha'^2}{\kappa^2}\int d^4x \sqrt{-g}\left(\frac{1}{48}I_1 + \frac{1}{24}G_3\right),
\end{equation}
where $I_1$ is the $\beta_2$ term in our case, and 
\begin{equation}
    G_3 = R_{\mu\nu}^{\ \ \ \alpha\beta}R_{\alpha\beta}^{\ \ \ \lambda\sigma}R_{\lambda\sigma}^{\ \ \ \mu\nu} -2R^{\mu}_{\ \ \alpha\nu\beta}R^{\alpha \lambda\beta\sigma}R_{\lambda\mu\sigma}^{\ \ \ \ \nu},
\end{equation}
is the well known cubic Gauss-Bonnet invariant, which arises from setting $\beta_1 = -2\beta_2$. We note that in the case considered there is no contribution at order $\alpha'$ on-shell, since the second Gauss-Bonnet invariant is a surface term unless non-minimally coupled to a dynamic dilaton \cite{Tseytlin:1986zz,Deser:1986xr}. See Ref. \cite{Huber:2019ugz} for an analysis from the on-shell amplitude perspective.
\newline\newline
To compute the $I_1$ correction, we take $\lambda = -\frac{2\alpha'^2}{\kappa^2}$ and then $\beta_1 = 0,~\beta_2 = \frac{1}{48}$ which gives
\begin{flalign}
    \Phi^{I_1} &= \frac{3(G\alpha')^2}{2}\sinh^2w\,\cos(\textbf{a}\cdot\nabla)\frac{m_B^2}{r^6},\\
    A^{I_1}_i &= -18(G\alpha')^2m_B^2\sinh^2w\,\sin(\textbf{a}\cdot\nabla) \frac{(\mathbf{a}\times\mathbf{r})_i}{r^8}.
\end{flalign}
Then to compute the $G_3$ contribution we take $\beta_1 = -2\beta_2 = -\frac{1}{24}$ to find 
\begin{flalign}
    \Phi^{G_3} &= 3(G\alpha')^2\cos(\textbf{a}\cdot\nabla)\frac{m_B^2}{r^6},\\
    A^{G_3}_i &= - 9(G\alpha')^2m_B^2\sin(\textbf{a}\cdot\nabla)\frac{(\mathbf{a}\times\mathbf{r})_i}{r^8}.
\end{flalign}
We can also readily derive the scattering angle in this case, which we find to be
\begin{equation}\label{key}
\theta_{\alpha'^2} = \frac{45\pi}{32}(GM\alpha')^2\sum_\pm\frac{1}{(b\pm a)^6}.
\end{equation}
\section{Discussion}
In this paper, we have derived a new black hole solution in cubic gravity using modern amplitude methods. In particular, we present the all-order (in spin) classical potential, to leading-order in the cubic coupling $\lambda$. Further, we have shown how the form of this potential allows for an interpretation of the on-shell avatar of the Newman-Janis algorithm, extending it to higher-derivative gravity. This is certainly good motivation to try and establish the precise algorithm that deforms coordinates in some particular coordinate system and allows one to derive the rotating solution directly from the static one. As an additional exercise, we broadened our discussion slightly to consider the leading-order corrections to the potential arising from the $\alpha'^2$ part of the low-energy effective action in string theory. We also note that taking the $a\longrightarrow 0$ limit reproduces the results that can be found elsewhere in the literature \cite{Bueno:2016lrh,Bueno:2016xff,Cano:2019,Burger:2017yod,Hennigar:2016gkm,Brandhuber:2019qpg,Emond:2019crr}, at least for those quantities that have already been computed.

It should be emphasised that deriving such a black hole solution using traditional geometric methods is a difficult endeavour, a fact which highlights the benefits of using modern amplitude techniques to understand gravitational phenomena. Indeed, \textit{this is a rare case in which it has been possible to derive a novel result via modern amplitude methods before it has been done so through the geometric approach}, in which the presence of cubic-order curvature terms makes the task almost intractable.

In addition to finding the black hole solution, we also present results for pertinent classical observables, namely the impulse and scattering angle. Given some reliable observational data, these quantities could be used to place a bound on the coupling $\lambda$, whose parameter space has been only partially constrained \cite{Poshteh:2018wqy} -- in the model presented, $\lambda$ can assume arbitrarily large or small values.

There are a number of interesting future directions that immediately present themselves as natural follow ups to this work. It is certainly plausible to consider a binary rotating black hole system cubic gravity, which would amount to including spin on all external particles in the scattering amplitudes and keeping the particle masses general. Furthermore, one could consider gravitational radiation in this setup, which would give rise to observational signatures that could, at least in principle, be detected by LIGO or future gravitational wave experiments, such a LISA.\newline

\textit{Addendum}: We note with interest that, simultaneously submitted to the arXiv along with this paper was Ref. \cite{Cano:2019ozf}, in which the authors study the extremal near-horizon geometry of rotating black holes in ECG. Interestingly, they note that a forthcoming publication will study rotating solutions to ECG in general, and we eagerly look forward to their results and the comparison to those we have derived using amplitude techniques.

\subsection{Acknowledgements}
The authors would like to thank Jeff Murugan for useful comments and discussions. DB \& NM are supported by the South African Research Chairs Initiative of the Department of Science and Technology and the National Research Foundation of South Africa. Any opinion, finding and conclusion or recommendation expressed in this material is that of the authors and the NRF does not accept any liability in this regard. WE is supported by the STFC under grant no. ST/P000703/1.

\bibliography{mainPRD}

\begin{thebibliography}{50}%
\makeatletter
\providecommand \@ifxundefined [1]{%
 \@ifx{#1\undefined}
}%
\providecommand \@ifnum [1]{%
 \ifnum #1\expandafter \@firstoftwo
 \else \expandafter \@secondoftwo
 \fi
}%
\providecommand \@ifx [1]{%
 \ifx #1\expandafter \@firstoftwo
 \else \expandafter \@secondoftwo
 \fi
}%
\providecommand \natexlab [1]{#1}%
\providecommand \enquote  [1]{``#1''}%
\providecommand \bibnamefont  [1]{#1}%
\providecommand \bibfnamefont [1]{#1}%
\providecommand \citenamefont [1]{#1}%
\providecommand \href@noop [0]{\@secondoftwo}%
\providecommand \href [0]{\begingroup \@sanitize@url \@href}%
\providecommand \@href[1]{\@@startlink{#1}\@@href}%
\providecommand \@@href[1]{\endgroup#1\@@endlink}%
\providecommand \@sanitize@url [0]{\catcode `\\12\catcode `\$12\catcode
  `\&12\catcode `\#12\catcode `\^12\catcode `\_12\catcode `\%12\relax}%
\providecommand \@@startlink[1]{}%
\providecommand \@@endlink[0]{}%
\providecommand \url  [0]{\begingroup\@sanitize@url \@url }%
\providecommand \@url [1]{\endgroup\@href {#1}{\urlprefix }}%
\providecommand \urlprefix  [0]{URL }%
\providecommand \Eprint [0]{\href }%
\providecommand \doibase [0]{http://dx.doi.org/}%
\providecommand \selectlanguage [0]{\@gobble}%
\providecommand \bibinfo  [0]{\@secondoftwo}%
\providecommand \bibfield  [0]{\@secondoftwo}%
\providecommand \translation [1]{[#1]}%
\providecommand \BibitemOpen [0]{}%
\providecommand \bibitemStop [0]{}%
\providecommand \bibitemNoStop [0]{.\EOS\space}%
\providecommand \EOS [0]{\spacefactor3000\relax}%
\providecommand \BibitemShut  [1]{\csname bibitem#1\endcsname}%
\let\auto@bib@innerbib\@empty
\bibitem [{\citenamefont {Abbott}\ \emph {et~al.}(2016)\citenamefont {Abbott}
  \emph {et~al.}}]{Abbott:2016blz}%
  \BibitemOpen
  \bibfield  {author} {\bibinfo {author} {\bibfnamefont {B.~P.}\ \bibnamefont
  {Abbott}} \emph {et~al.} (\bibinfo {collaboration} {LIGO Scientific,
  Virgo}),\ }\href {\doibase 10.1103/PhysRevLett.116.061102} {\bibfield
  {journal} {\bibinfo  {journal} {Phys. Rev. Lett.}\ }\textbf {\bibinfo
  {volume} {116}},\ \bibinfo {pages} {061102} (\bibinfo {year} {2016})},\
  \Eprint {http://arxiv.org/abs/1602.03837} {arXiv:1602.03837 [gr-qc]}
  \BibitemShut {NoStop}%
\bibitem [{\citenamefont {Abbott}\ \emph {et~al.}(2017)\citenamefont {Abbott}
  \emph {et~al.}}]{TheLIGOScientific:2017qsa}%
  \BibitemOpen
  \bibfield  {author} {\bibinfo {author} {\bibfnamefont {B.~P.}\ \bibnamefont
  {Abbott}} \emph {et~al.} (\bibinfo {collaboration} {LIGO Scientific,
  Virgo}),\ }\href {\doibase 10.1103/PhysRevLett.119.161101} {\bibfield
  {journal} {\bibinfo  {journal} {Phys. Rev. Lett.}\ }\textbf {\bibinfo
  {volume} {119}},\ \bibinfo {pages} {161101} (\bibinfo {year} {2017})},\
  \Eprint {http://arxiv.org/abs/1710.05832} {arXiv:1710.05832 [gr-qc]}
  \BibitemShut {NoStop}%
\bibitem [{\citenamefont {Emond}\ and\ \citenamefont
  {Moynihan}(2019)}]{Emond:2019crr}%
  \BibitemOpen
  \bibfield  {author} {\bibinfo {author} {\bibfnamefont {W.~T.}\ \bibnamefont
  {Emond}}\ and\ \bibinfo {author} {\bibfnamefont {N.}~\bibnamefont
  {Moynihan}},\ }\href@noop {} {\  (\bibinfo {year} {2019})},\ \Eprint
  {http://arxiv.org/abs/1905.08213} {arXiv:1905.08213 [hep-th]} \BibitemShut
  {NoStop}%
\bibitem [{\citenamefont {Holstein}\ and\ \citenamefont
  {Ross}(2008)}]{Holstein:2008sx}%
  \BibitemOpen
  \bibfield  {author} {\bibinfo {author} {\bibfnamefont {B.~R.}\ \bibnamefont
  {Holstein}}\ and\ \bibinfo {author} {\bibfnamefont {A.}~\bibnamefont
  {Ross}},\ }\href@noop {} {\  (\bibinfo {year} {2008})},\ \Eprint
  {http://arxiv.org/abs/0802.0716} {arXiv:0802.0716 [hep-ph]} \BibitemShut
  {NoStop}%
\bibitem [{\citenamefont {Neill}\ and\ \citenamefont
  {Rothstein}(2013)}]{Neill:2013wsa}%
  \BibitemOpen
  \bibfield  {author} {\bibinfo {author} {\bibfnamefont {D.}~\bibnamefont
  {Neill}}\ and\ \bibinfo {author} {\bibfnamefont {I.~Z.}\ \bibnamefont
  {Rothstein}},\ }\href {\doibase 10.1016/j.nuclphysb.2013.09.007} {\bibfield
  {journal} {\bibinfo  {journal} {Nucl. Phys.}\ }\textbf {\bibinfo {volume}
  {B877}},\ \bibinfo {pages} {177} (\bibinfo {year} {2013})},\ \Eprint
  {http://arxiv.org/abs/1304.7263} {arXiv:1304.7263 [hep-th]} \BibitemShut
  {NoStop}%
\bibitem [{\citenamefont {Vaidya}(2015)}]{Vaidya:2014kza}%
  \BibitemOpen
  \bibfield  {author} {\bibinfo {author} {\bibfnamefont {V.}~\bibnamefont
  {Vaidya}},\ }\href {\doibase 10.1103/PhysRevD.91.024017} {\bibfield
  {journal} {\bibinfo  {journal} {Phys. Rev.}\ }\textbf {\bibinfo {volume}
  {D91}},\ \bibinfo {pages} {024017} (\bibinfo {year} {2015})},\ \Eprint
  {http://arxiv.org/abs/1410.5348} {arXiv:1410.5348 [hep-th]} \BibitemShut
  {NoStop}%
\bibitem [{\citenamefont {Damour}(2016)}]{Damour:2016gwp}%
  \BibitemOpen
  \bibfield  {author} {\bibinfo {author} {\bibfnamefont {T.}~\bibnamefont
  {Damour}},\ }\href {\doibase 10.1103/PhysRevD.94.104015} {\bibfield
  {journal} {\bibinfo  {journal} {Phys. Rev.}\ }\textbf {\bibinfo {volume}
  {D94}},\ \bibinfo {pages} {104015} (\bibinfo {year} {2016})},\ \Eprint
  {http://arxiv.org/abs/1609.00354} {arXiv:1609.00354 [gr-qc]} \BibitemShut
  {NoStop}%
\bibitem [{\citenamefont {Cachazo}\ and\ \citenamefont
  {Guevara}(2017)}]{Cachazo:2017jef}%
  \BibitemOpen
  \bibfield  {author} {\bibinfo {author} {\bibfnamefont {F.}~\bibnamefont
  {Cachazo}}\ and\ \bibinfo {author} {\bibfnamefont {A.}~\bibnamefont
  {Guevara}},\ }\href@noop {} {\  (\bibinfo {year} {2017})},\ \Eprint
  {http://arxiv.org/abs/1705.10262} {arXiv:1705.10262 [hep-th]} \BibitemShut
  {NoStop}%
\bibitem [{\citenamefont {Cheung}\ \emph {et~al.}(2018)\citenamefont {Cheung},
  \citenamefont {Rothstein},\ and\ \citenamefont {Solon}}]{Cheung:2018wkq}%
  \BibitemOpen
  \bibfield  {author} {\bibinfo {author} {\bibfnamefont {C.}~\bibnamefont
  {Cheung}}, \bibinfo {author} {\bibfnamefont {I.~Z.}\ \bibnamefont
  {Rothstein}}, \ and\ \bibinfo {author} {\bibfnamefont {M.~P.}\ \bibnamefont
  {Solon}},\ }\href {\doibase 10.1103/PhysRevLett.121.251101} {\bibfield
  {journal} {\bibinfo  {journal} {Phys. Rev. Lett.}\ }\textbf {\bibinfo
  {volume} {121}},\ \bibinfo {pages} {251101} (\bibinfo {year} {2018})},\
  \Eprint {http://arxiv.org/abs/1808.02489} {arXiv:1808.02489 [hep-th]}
  \BibitemShut {NoStop}%
\bibitem [{\citenamefont {Carballo-Rubio}\ \emph {et~al.}(2018)\citenamefont
  {Carballo-Rubio}, \citenamefont {Di~Filippo},\ and\ \citenamefont
  {Moynihan}}]{Carballo-Rubio:2018bmu}%
  \BibitemOpen
  \bibfield  {author} {\bibinfo {author} {\bibfnamefont {R.}~\bibnamefont
  {Carballo-Rubio}}, \bibinfo {author} {\bibfnamefont {F.}~\bibnamefont
  {Di~Filippo}}, \ and\ \bibinfo {author} {\bibfnamefont {N.}~\bibnamefont
  {Moynihan}},\ }\href@noop {} {\  (\bibinfo {year} {2018})},\ \Eprint
  {http://arxiv.org/abs/1811.08192} {arXiv:1811.08192 [hep-th]} \BibitemShut
  {NoStop}%
\bibitem [{\citenamefont {Guevara}\ \emph
  {et~al.}(2019{\natexlab{a}})\citenamefont {Guevara}, \citenamefont
  {Ochirov},\ and\ \citenamefont {Vines}}]{Guevara:2018wpp}%
  \BibitemOpen
  \bibfield  {author} {\bibinfo {author} {\bibfnamefont {A.}~\bibnamefont
  {Guevara}}, \bibinfo {author} {\bibfnamefont {A.}~\bibnamefont {Ochirov}}, \
  and\ \bibinfo {author} {\bibfnamefont {J.}~\bibnamefont {Vines}},\ }\href
  {\doibase 10.1007/JHEP09(2019)056} {\bibfield  {journal} {\bibinfo  {journal}
  {JHEP}\ }\textbf {\bibinfo {volume} {09}},\ \bibinfo {pages} {056} (\bibinfo
  {year} {2019}{\natexlab{a}})},\ \Eprint {http://arxiv.org/abs/1812.06895}
  {arXiv:1812.06895 [hep-th]} \BibitemShut {NoStop}%
\bibitem [{\citenamefont {Chung}\ \emph
  {et~al.}(2019{\natexlab{a}})\citenamefont {Chung}, \citenamefont {Huang},
  \citenamefont {Kim},\ and\ \citenamefont {Lee}}]{Chung:2018kqs}%
  \BibitemOpen
  \bibfield  {author} {\bibinfo {author} {\bibfnamefont {M.-Z.}\ \bibnamefont
  {Chung}}, \bibinfo {author} {\bibfnamefont {Y.-T.}\ \bibnamefont {Huang}},
  \bibinfo {author} {\bibfnamefont {J.-W.}\ \bibnamefont {Kim}}, \ and\
  \bibinfo {author} {\bibfnamefont {S.}~\bibnamefont {Lee}},\ }\href {\doibase
  10.1007/JHEP04(2019)156} {\bibfield  {journal} {\bibinfo  {journal} {JHEP}\
  }\textbf {\bibinfo {volume} {04}},\ \bibinfo {pages} {156} (\bibinfo {year}
  {2019}{\natexlab{a}})},\ \Eprint {http://arxiv.org/abs/1812.08752}
  {arXiv:1812.08752 [hep-th]} \BibitemShut {NoStop}%
\bibitem [{\citenamefont {Caron-Huot}\ and\ \citenamefont
  {Zahraee}(2019)}]{Caron-Huot:2018ape}%
  \BibitemOpen
  \bibfield  {author} {\bibinfo {author} {\bibfnamefont {S.}~\bibnamefont
  {Caron-Huot}}\ and\ \bibinfo {author} {\bibfnamefont {Z.}~\bibnamefont
  {Zahraee}},\ }\href {\doibase 10.1007/JHEP07(2019)179} {\bibfield  {journal}
  {\bibinfo  {journal} {JHEP}\ }\textbf {\bibinfo {volume} {07}},\ \bibinfo
  {pages} {179} (\bibinfo {year} {2019})},\ \Eprint
  {http://arxiv.org/abs/1810.04694} {arXiv:1810.04694 [hep-th]} \BibitemShut
  {NoStop}%
\bibitem [{\citenamefont {Brandhuber}\ and\ \citenamefont
  {Travaglini}(2019)}]{Brandhuber:2019qpg}%
  \BibitemOpen
  \bibfield  {author} {\bibinfo {author} {\bibfnamefont {A.}~\bibnamefont
  {Brandhuber}}\ and\ \bibinfo {author} {\bibfnamefont {G.}~\bibnamefont
  {Travaglini}},\ }\href@noop {} {\  (\bibinfo {year} {2019})},\ \Eprint
  {http://arxiv.org/abs/1905.05657} {arXiv:1905.05657 [hep-th]} \BibitemShut
  {NoStop}%
\bibitem [{\citenamefont {Cristofoli}(2019)}]{Cristofoli:2019ewu}%
  \BibitemOpen
  \bibfield  {author} {\bibinfo {author} {\bibfnamefont {A.}~\bibnamefont
  {Cristofoli}},\ }\href@noop {} {\  (\bibinfo {year} {2019})},\ \Eprint
  {http://arxiv.org/abs/1906.05209} {arXiv:1906.05209 [hep-th]} \BibitemShut
  {NoStop}%
\bibitem [{\citenamefont {Chung}\ \emph
  {et~al.}(2019{\natexlab{b}})\citenamefont {Chung}, \citenamefont {Huang},\
  and\ \citenamefont {Kim}}]{Chung:2019duq}%
  \BibitemOpen
  \bibfield  {author} {\bibinfo {author} {\bibfnamefont {M.-Z.}\ \bibnamefont
  {Chung}}, \bibinfo {author} {\bibfnamefont {Y.-T.}\ \bibnamefont {Huang}}, \
  and\ \bibinfo {author} {\bibfnamefont {J.-W.}\ \bibnamefont {Kim}},\
  }\href@noop {} {\  (\bibinfo {year} {2019}{\natexlab{b}})},\ \Eprint
  {http://arxiv.org/abs/1908.08463} {arXiv:1908.08463 [hep-th]} \BibitemShut
  {NoStop}%
\bibitem [{\citenamefont {Guevara}\ \emph
  {et~al.}(2019{\natexlab{b}})\citenamefont {Guevara}, \citenamefont
  {Ochirov},\ and\ \citenamefont {Vines}}]{Guevara:2019fsj}%
  \BibitemOpen
  \bibfield  {author} {\bibinfo {author} {\bibfnamefont {A.}~\bibnamefont
  {Guevara}}, \bibinfo {author} {\bibfnamefont {A.}~\bibnamefont {Ochirov}}, \
  and\ \bibinfo {author} {\bibfnamefont {J.}~\bibnamefont {Vines}},\
  }\href@noop {} {\  (\bibinfo {year} {2019}{\natexlab{b}})},\ \Eprint
  {http://arxiv.org/abs/1906.10071} {arXiv:1906.10071 [hep-th]} \BibitemShut
  {NoStop}%
\bibitem [{\citenamefont {Bern}\ \emph {et~al.}(2019)\citenamefont {Bern},
  \citenamefont {Cheung}, \citenamefont {Roiban}, \citenamefont {Shen},
  \citenamefont {Solon},\ and\ \citenamefont {Zeng}}]{Bern:2019crd}%
  \BibitemOpen
  \bibfield  {author} {\bibinfo {author} {\bibfnamefont {Z.}~\bibnamefont
  {Bern}}, \bibinfo {author} {\bibfnamefont {C.}~\bibnamefont {Cheung}},
  \bibinfo {author} {\bibfnamefont {R.}~\bibnamefont {Roiban}}, \bibinfo
  {author} {\bibfnamefont {C.-H.}\ \bibnamefont {Shen}}, \bibinfo {author}
  {\bibfnamefont {M.~P.}\ \bibnamefont {Solon}}, \ and\ \bibinfo {author}
  {\bibfnamefont {M.}~\bibnamefont {Zeng}},\ }\href@noop {} {\  (\bibinfo
  {year} {2019})},\ \Eprint {http://arxiv.org/abs/1908.01493} {arXiv:1908.01493
  [hep-th]} \BibitemShut {NoStop}%
\bibitem [{\citenamefont {Arkani-Hamed}\ \emph {et~al.}(2019)\citenamefont
  {Arkani-Hamed}, \citenamefont {Huang},\ and\ \citenamefont
  {O'Connell}}]{Arkani-Hamed:2019ymq}%
  \BibitemOpen
  \bibfield  {author} {\bibinfo {author} {\bibfnamefont {N.}~\bibnamefont
  {Arkani-Hamed}}, \bibinfo {author} {\bibfnamefont {Y.-t.}\ \bibnamefont
  {Huang}}, \ and\ \bibinfo {author} {\bibfnamefont {D.}~\bibnamefont
  {O'Connell}},\ }\href@noop {} {\  (\bibinfo {year} {2019})},\ \Eprint
  {http://arxiv.org/abs/1906.10100} {arXiv:1906.10100 [hep-th]} \BibitemShut
  {NoStop}%
\bibitem [{\citenamefont {Moynihan}(2019)}]{Moynihan:2019bor}%
  \BibitemOpen
  \bibfield  {author} {\bibinfo {author} {\bibfnamefont {N.}~\bibnamefont
  {Moynihan}},\ }\href@noop {} {\  (\bibinfo {year} {2019})},\ \Eprint
  {http://arxiv.org/abs/1909.05217} {arXiv:1909.05217 [hep-th]} \BibitemShut
  {NoStop}%
\bibitem [{\citenamefont {Siemonsen}\ and\ \citenamefont
  {Vines}(2019)}]{Siemonsen:2019dsu}%
  \BibitemOpen
  \bibfield  {author} {\bibinfo {author} {\bibfnamefont {N.}~\bibnamefont
  {Siemonsen}}\ and\ \bibinfo {author} {\bibfnamefont {J.}~\bibnamefont
  {Vines}},\ }\href@noop {} {\  (\bibinfo {year} {2019})},\ \Eprint
  {http://arxiv.org/abs/1909.07361} {arXiv:1909.07361 [gr-qc]} \BibitemShut
  {NoStop}%
\bibitem [{\citenamefont {Kälin}\ and\ \citenamefont
  {Porto}(2019)}]{Kalin:2019rwq}%
  \BibitemOpen
  \bibfield  {author} {\bibinfo {author} {\bibfnamefont {G.}~\bibnamefont
  {Kälin}}\ and\ \bibinfo {author} {\bibfnamefont {R.~A.}\ \bibnamefont
  {Porto}},\ }\href@noop {} {\  (\bibinfo {year} {2019})},\ \Eprint
  {http://arxiv.org/abs/1910.03008} {arXiv:1910.03008 [hep-th]} \BibitemShut
  {NoStop}%
\bibitem [{\citenamefont {Porto}(2006)}]{Porto:2005ac}%
  \BibitemOpen
  \bibfield  {author} {\bibinfo {author} {\bibfnamefont {R.~A.}\ \bibnamefont
  {Porto}},\ }\href {\doibase 10.1103/PhysRevD.73.104031} {\bibfield  {journal}
  {\bibinfo  {journal} {Phys. Rev.}\ }\textbf {\bibinfo {volume} {D73}},\
  \bibinfo {pages} {104031} (\bibinfo {year} {2006})},\ \Eprint
  {http://arxiv.org/abs/gr-qc/0511061} {arXiv:gr-qc/0511061 [gr-qc]}
  \BibitemShut {NoStop}%
\bibitem [{\citenamefont {Levi}(2010)}]{Levi:2010zu}%
  \BibitemOpen
  \bibfield  {author} {\bibinfo {author} {\bibfnamefont {M.}~\bibnamefont
  {Levi}},\ }\href {\doibase 10.1103/PhysRevD.82.104004} {\bibfield  {journal}
  {\bibinfo  {journal} {Phys. Rev.}\ }\textbf {\bibinfo {volume} {D82}},\
  \bibinfo {pages} {104004} (\bibinfo {year} {2010})},\ \Eprint
  {http://arxiv.org/abs/1006.4139} {arXiv:1006.4139 [gr-qc]} \BibitemShut
  {NoStop}%
\bibitem [{\citenamefont {Levi}\ and\ \citenamefont
  {Steinhoff}(2015{\natexlab{a}})}]{Levi:2014gsa}%
  \BibitemOpen
  \bibfield  {author} {\bibinfo {author} {\bibfnamefont {M.}~\bibnamefont
  {Levi}}\ and\ \bibinfo {author} {\bibfnamefont {J.}~\bibnamefont
  {Steinhoff}},\ }\href {\doibase 10.1007/JHEP06(2015)059} {\bibfield
  {journal} {\bibinfo  {journal} {JHEP}\ }\textbf {\bibinfo {volume} {06}},\
  \bibinfo {pages} {059} (\bibinfo {year} {2015}{\natexlab{a}})},\ \Eprint
  {http://arxiv.org/abs/1410.2601} {arXiv:1410.2601 [gr-qc]} \BibitemShut
  {NoStop}%
\bibitem [{\citenamefont {Levi}\ and\ \citenamefont
  {Steinhoff}(2015{\natexlab{b}})}]{Levi:2015msa}%
  \BibitemOpen
  \bibfield  {author} {\bibinfo {author} {\bibfnamefont {M.}~\bibnamefont
  {Levi}}\ and\ \bibinfo {author} {\bibfnamefont {J.}~\bibnamefont
  {Steinhoff}},\ }\href {\doibase 10.1007/JHEP09(2015)219} {\bibfield
  {journal} {\bibinfo  {journal} {JHEP}\ }\textbf {\bibinfo {volume} {09}},\
  \bibinfo {pages} {219} (\bibinfo {year} {2015}{\natexlab{b}})},\ \Eprint
  {http://arxiv.org/abs/1501.04956} {arXiv:1501.04956 [gr-qc]} \BibitemShut
  {NoStop}%
\bibitem [{\citenamefont {Vines}\ and\ \citenamefont
  {Steinhoff}(2018)}]{Vines:2016qwa}%
  \BibitemOpen
  \bibfield  {author} {\bibinfo {author} {\bibfnamefont {J.}~\bibnamefont
  {Vines}}\ and\ \bibinfo {author} {\bibfnamefont {J.}~\bibnamefont
  {Steinhoff}},\ }\href {\doibase 10.1103/PhysRevD.97.064010} {\bibfield
  {journal} {\bibinfo  {journal} {Phys. Rev.}\ }\textbf {\bibinfo {volume}
  {D97}},\ \bibinfo {pages} {064010} (\bibinfo {year} {2018})},\ \Eprint
  {http://arxiv.org/abs/1606.08832} {arXiv:1606.08832 [gr-qc]} \BibitemShut
  {NoStop}%
\bibitem [{\citenamefont {Vines}(2018)}]{Vines:2017hyw}%
  \BibitemOpen
  \bibfield  {author} {\bibinfo {author} {\bibfnamefont {J.}~\bibnamefont
  {Vines}},\ }\href {\doibase 10.1088/1361-6382/aaa3a8} {\bibfield  {journal}
  {\bibinfo  {journal} {Class. Quant. Grav.}\ }\textbf {\bibinfo {volume}
  {35}},\ \bibinfo {pages} {084002} (\bibinfo {year} {2018})},\ \Eprint
  {http://arxiv.org/abs/1709.06016} {arXiv:1709.06016 [gr-qc]} \BibitemShut
  {NoStop}%
\bibitem [{\citenamefont {Vines}\ \emph {et~al.}(2019)\citenamefont {Vines},
  \citenamefont {Steinhoff},\ and\ \citenamefont {Buonanno}}]{Vines:2018gqi}%
  \BibitemOpen
  \bibfield  {author} {\bibinfo {author} {\bibfnamefont {J.}~\bibnamefont
  {Vines}}, \bibinfo {author} {\bibfnamefont {J.}~\bibnamefont {Steinhoff}}, \
  and\ \bibinfo {author} {\bibfnamefont {A.}~\bibnamefont {Buonanno}},\ }\href
  {\doibase 10.1103/PhysRevD.99.064054} {\bibfield  {journal} {\bibinfo
  {journal} {Phys. Rev.}\ }\textbf {\bibinfo {volume} {D99}},\ \bibinfo {pages}
  {064054} (\bibinfo {year} {2019})},\ \Eprint
  {http://arxiv.org/abs/1812.00956} {arXiv:1812.00956 [gr-qc]} \BibitemShut
  {NoStop}%
\bibitem [{\citenamefont {Levi}(2018)}]{Levi:2018nxp}%
  \BibitemOpen
  \bibfield  {author} {\bibinfo {author} {\bibfnamefont {M.}~\bibnamefont
  {Levi}},\ }\href@noop {} {\  (\bibinfo {year} {2018})},\ \Eprint
  {http://arxiv.org/abs/1807.01699} {arXiv:1807.01699 [hep-th]} \BibitemShut
  {NoStop}%
\bibitem [{\citenamefont {Guevara}(2019)}]{Guevara:2017csg}%
  \BibitemOpen
  \bibfield  {author} {\bibinfo {author} {\bibfnamefont {A.}~\bibnamefont
  {Guevara}},\ }\href {\doibase 10.1007/JHEP04(2019)033} {\bibfield  {journal}
  {\bibinfo  {journal} {JHEP}\ }\textbf {\bibinfo {volume} {04}},\ \bibinfo
  {pages} {033} (\bibinfo {year} {2019})},\ \Eprint
  {http://arxiv.org/abs/1706.02314} {arXiv:1706.02314 [hep-th]} \BibitemShut
  {NoStop}%
\bibitem [{\citenamefont {Sisman}\ \emph {et~al.}(2011)\citenamefont {Sisman},
  \citenamefont {Gullu},\ and\ \citenamefont {Tekin}}]{Sisman:2011gz}%
  \BibitemOpen
  \bibfield  {author} {\bibinfo {author} {\bibfnamefont {T.~C.}\ \bibnamefont
  {Sisman}}, \bibinfo {author} {\bibfnamefont {I.}~\bibnamefont {Gullu}}, \
  and\ \bibinfo {author} {\bibfnamefont {B.}~\bibnamefont {Tekin}},\ }\href
  {\doibase 10.1088/0264-9381/28/19/195004} {\bibfield  {journal} {\bibinfo
  {journal} {Class. Quant. Grav.}\ }\textbf {\bibinfo {volume} {28}},\ \bibinfo
  {pages} {195004} (\bibinfo {year} {2011})},\ \Eprint
  {http://arxiv.org/abs/1103.2307} {arXiv:1103.2307 [hep-th]} \BibitemShut
  {NoStop}%
\bibitem [{\citenamefont {Bueno}\ and\ \citenamefont
  {Cano}(2016{\natexlab{a}})}]{Bueno:2016lrh}%
  \BibitemOpen
  \bibfield  {author} {\bibinfo {author} {\bibfnamefont {P.}~\bibnamefont
  {Bueno}}\ and\ \bibinfo {author} {\bibfnamefont {P.~A.}\ \bibnamefont
  {Cano}},\ }\href {\doibase 10.1103/PhysRevD.94.124051} {\bibfield  {journal}
  {\bibinfo  {journal} {Phys. Rev.}\ }\textbf {\bibinfo {volume} {D94}},\
  \bibinfo {pages} {124051} (\bibinfo {year} {2016}{\natexlab{a}})},\ \Eprint
  {http://arxiv.org/abs/1610.08019} {arXiv:1610.08019 [hep-th]} \BibitemShut
  {NoStop}%
\bibitem [{\citenamefont {Bueno}\ and\ \citenamefont
  {Cano}(2016{\natexlab{b}})}]{Bueno:2016xff}%
  \BibitemOpen
  \bibfield  {author} {\bibinfo {author} {\bibfnamefont {P.}~\bibnamefont
  {Bueno}}\ and\ \bibinfo {author} {\bibfnamefont {P.~A.}\ \bibnamefont
  {Cano}},\ }\href {\doibase 10.1103/PhysRevD.94.104005} {\bibfield  {journal}
  {\bibinfo  {journal} {Phys. Rev.}\ }\textbf {\bibinfo {volume} {D94}},\
  \bibinfo {pages} {104005} (\bibinfo {year} {2016}{\natexlab{b}})},\ \Eprint
  {http://arxiv.org/abs/1607.06463} {arXiv:1607.06463 [hep-th]} \BibitemShut
  {NoStop}%
\bibitem [{\citenamefont {Hennigar}\ and\ \citenamefont
  {Mann}(2017)}]{Hennigar:2016gkm}%
  \BibitemOpen
  \bibfield  {author} {\bibinfo {author} {\bibfnamefont {R.~A.}\ \bibnamefont
  {Hennigar}}\ and\ \bibinfo {author} {\bibfnamefont {R.~B.}\ \bibnamefont
  {Mann}},\ }\href {\doibase 10.1103/PhysRevD.95.064055} {\bibfield  {journal}
  {\bibinfo  {journal} {Phys. Rev.}\ }\textbf {\bibinfo {volume} {D95}},\
  \bibinfo {pages} {064055} (\bibinfo {year} {2017})},\ \Eprint
  {http://arxiv.org/abs/1610.06675} {arXiv:1610.06675 [hep-th]} \BibitemShut
  {NoStop}%
\bibitem [{\citenamefont {Hennigar}\ \emph {et~al.}(2017)\citenamefont
  {Hennigar}, \citenamefont {Kubizňák},\ and\ \citenamefont
  {Mann}}]{Hennigar:2017ego}%
  \BibitemOpen
  \bibfield  {author} {\bibinfo {author} {\bibfnamefont {R.~A.}\ \bibnamefont
  {Hennigar}}, \bibinfo {author} {\bibfnamefont {D.}~\bibnamefont
  {Kubizňák}}, \ and\ \bibinfo {author} {\bibfnamefont {R.~B.}\ \bibnamefont
  {Mann}},\ }\href {\doibase 10.1103/PhysRevD.95.104042} {\bibfield  {journal}
  {\bibinfo  {journal} {Phys. Rev.}\ }\textbf {\bibinfo {volume} {D95}},\
  \bibinfo {pages} {104042} (\bibinfo {year} {2017})},\ \Eprint
  {http://arxiv.org/abs/1703.01631} {arXiv:1703.01631 [hep-th]} \BibitemShut
  {NoStop}%
\bibitem [{\citenamefont {Poshteh}\ and\ \citenamefont
  {Mann}(2019)}]{Poshteh:2018wqy}%
  \BibitemOpen
  \bibfield  {author} {\bibinfo {author} {\bibfnamefont {M.~B.~J.}\
  \bibnamefont {Poshteh}}\ and\ \bibinfo {author} {\bibfnamefont {R.~B.}\
  \bibnamefont {Mann}},\ }\href {\doibase 10.1103/PhysRevD.99.024035}
  {\bibfield  {journal} {\bibinfo  {journal} {Phys. Rev.}\ }\textbf {\bibinfo
  {volume} {D99}},\ \bibinfo {pages} {024035} (\bibinfo {year} {2019})},\
  \Eprint {http://arxiv.org/abs/1810.10657} {arXiv:1810.10657 [gr-qc]}
  \BibitemShut {NoStop}%
\bibitem [{\citenamefont {Mir}\ and\ \citenamefont {Mann}(2019)}]{Mir:2019rik}%
  \BibitemOpen
  \bibfield  {author} {\bibinfo {author} {\bibfnamefont {M.}~\bibnamefont
  {Mir}}\ and\ \bibinfo {author} {\bibfnamefont {R.~B.}\ \bibnamefont {Mann}},\
  }\href {\doibase 10.1007/JHEP07(2019)012} {\bibfield  {journal} {\bibinfo
  {journal} {JHEP}\ }\textbf {\bibinfo {volume} {07}},\ \bibinfo {pages} {012}
  (\bibinfo {year} {2019})},\ \Eprint {http://arxiv.org/abs/1902.10906}
  {arXiv:1902.10906 [hep-th]} \BibitemShut {NoStop}%
\bibitem [{\citenamefont {Metsaev}\ and\ \citenamefont
  {Tseytlin}(1987)}]{Metsaev:1986yb}%
  \BibitemOpen
  \bibfield  {author} {\bibinfo {author} {\bibfnamefont {R.~R.}\ \bibnamefont
  {Metsaev}}\ and\ \bibinfo {author} {\bibfnamefont {A.~A.}\ \bibnamefont
  {Tseytlin}},\ }\href {\doibase 10.1016/0370-2693(87)91527-9} {\bibfield
  {journal} {\bibinfo  {journal} {Phys. Lett.}\ }\textbf {\bibinfo {volume}
  {B185}},\ \bibinfo {pages} {52} (\bibinfo {year} {1987})}\BibitemShut
  {NoStop}%
\bibitem [{\citenamefont {Arkani-Hamed}\ \emph {et~al.}(2017)\citenamefont
  {Arkani-Hamed}, \citenamefont {Huang},\ and\ \citenamefont
  {Huang}}]{Arkani-Hamed:2017jhn}%
  \BibitemOpen
  \bibfield  {author} {\bibinfo {author} {\bibfnamefont {N.}~\bibnamefont
  {Arkani-Hamed}}, \bibinfo {author} {\bibfnamefont {T.-C.}\ \bibnamefont
  {Huang}}, \ and\ \bibinfo {author} {\bibfnamefont {Y.-t.}\ \bibnamefont
  {Huang}},\ }\href@noop {} {\  (\bibinfo {year} {2017})},\ \Eprint
  {http://arxiv.org/abs/1709.04891} {arXiv:1709.04891 [hep-th]} \BibitemShut
  {NoStop}%
\bibitem [{\citenamefont {Maybee}\ \emph {et~al.}(2019)\citenamefont {Maybee},
  \citenamefont {O'Connell},\ and\ \citenamefont {Vines}}]{Maybee:2019jus}%
  \BibitemOpen
  \bibfield  {author} {\bibinfo {author} {\bibfnamefont {B.}~\bibnamefont
  {Maybee}}, \bibinfo {author} {\bibfnamefont {D.}~\bibnamefont {O'Connell}}, \
  and\ \bibinfo {author} {\bibfnamefont {J.}~\bibnamefont {Vines}},\
  }\href@noop {} {\  (\bibinfo {year} {2019})},\ \Eprint
  {http://arxiv.org/abs/1906.09260} {arXiv:1906.09260 [hep-th]} \BibitemShut
  {NoStop}%
\bibitem [{\citenamefont {Kosower}\ \emph {et~al.}(2019)\citenamefont
  {Kosower}, \citenamefont {Maybee},\ and\ \citenamefont
  {O'Connell}}]{Kosower:2018adc}%
  \BibitemOpen
  \bibfield  {author} {\bibinfo {author} {\bibfnamefont {D.~A.}\ \bibnamefont
  {Kosower}}, \bibinfo {author} {\bibfnamefont {B.}~\bibnamefont {Maybee}}, \
  and\ \bibinfo {author} {\bibfnamefont {D.}~\bibnamefont {O'Connell}},\ }\href
  {\doibase 10.1007/JHEP02(2019)137} {\bibfield  {journal} {\bibinfo  {journal}
  {JHEP}\ }\textbf {\bibinfo {volume} {02}},\ \bibinfo {pages} {137} (\bibinfo
  {year} {2019})},\ \Eprint {http://arxiv.org/abs/1811.10950} {arXiv:1811.10950
  [hep-th]} \BibitemShut {NoStop}%
\bibitem [{\citenamefont {Bjerrum-Bohr}\ \emph {et~al.}(2018)\citenamefont
  {Bjerrum-Bohr}, \citenamefont {Damgaard}, \citenamefont {Festuccia},
  \citenamefont {Plant\`e},\ and\ \citenamefont
  {Vanhove}}]{Bjerrum-Bohr:2018xdl}%
  \BibitemOpen
  \bibfield  {author} {\bibinfo {author} {\bibfnamefont {N.~E.~J.}\
  \bibnamefont {Bjerrum-Bohr}}, \bibinfo {author} {\bibfnamefont {P.~H.}\
  \bibnamefont {Damgaard}}, \bibinfo {author} {\bibfnamefont {G.}~\bibnamefont
  {Festuccia}}, \bibinfo {author} {\bibfnamefont {L.}~\bibnamefont {Plant\`e}},
  \ and\ \bibinfo {author} {\bibfnamefont {P.}~\bibnamefont {Vanhove}},\ }\href
  {\doibase 10.1103/PhysRevLett.121.171601} {\bibfield  {journal} {\bibinfo
  {journal} {Phys. Rev. Lett.}\ }\textbf {\bibinfo {volume} {121}},\ \bibinfo
  {pages} {171601} (\bibinfo {year} {2018})},\ \Eprint
  {http://arxiv.org/abs/1806.04920} {arXiv:1806.04920 [hep-th]} \BibitemShut
  {NoStop}%
\bibitem [{\citenamefont {Bjerrum-Bohr}\ \emph {et~al.}(2019)\citenamefont
  {Bjerrum-Bohr}, \citenamefont {Cristofoli},\ and\ \citenamefont
  {Damgaard}}]{Bjerrum-Bohr:2019kec}%
  \BibitemOpen
  \bibfield  {author} {\bibinfo {author} {\bibfnamefont {N.~E.~J.}\
  \bibnamefont {Bjerrum-Bohr}}, \bibinfo {author} {\bibfnamefont
  {A.}~\bibnamefont {Cristofoli}}, \ and\ \bibinfo {author} {\bibfnamefont
  {P.~H.}\ \bibnamefont {Damgaard}},\ }\href@noop {} {\  (\bibinfo {year}
  {2019})},\ \Eprint {http://arxiv.org/abs/1910.09366} {arXiv:1910.09366
  [hep-th]} \BibitemShut {NoStop}%
\bibitem [{\citenamefont {Tseytlin}(1986)}]{Tseytlin:1986zz}%
  \BibitemOpen
  \bibfield  {author} {\bibinfo {author} {\bibfnamefont {A.~A.}\ \bibnamefont
  {Tseytlin}},\ }\href {\doibase 10.1016/0370-2693(86)90930-5} {\bibfield
  {journal} {\bibinfo  {journal} {Phys. Lett.}\ }\textbf {\bibinfo {volume}
  {B176}},\ \bibinfo {pages} {92} (\bibinfo {year} {1986})}\BibitemShut
  {NoStop}%
\bibitem [{\citenamefont {Deser}\ and\ \citenamefont
  {Redlich}(1986)}]{Deser:1986xr}%
  \BibitemOpen
  \bibfield  {author} {\bibinfo {author} {\bibfnamefont {S.}~\bibnamefont
  {Deser}}\ and\ \bibinfo {author} {\bibfnamefont {A.~N.}\ \bibnamefont
  {Redlich}},\ }\href {\doibase 10.1016/0370-2693(86)90177-2} {\bibfield
  {journal} {\bibinfo  {journal} {Phys. Lett.}\ }\textbf {\bibinfo {volume}
  {B176}},\ \bibinfo {pages} {350} (\bibinfo {year} {1986})},\ \bibinfo {note}
  {[Erratum: Phys. Lett.B186,461(1987)]}\BibitemShut {NoStop}%
\bibitem [{\citenamefont {Accettulli~Huber}\ \emph {et~al.}(2019)\citenamefont
  {Accettulli~Huber}, \citenamefont {Brandhuber}, \citenamefont {De~Angelis},\
  and\ \citenamefont {Travaglini}}]{Huber:2019ugz}%
  \BibitemOpen
  \bibfield  {author} {\bibinfo {author} {\bibfnamefont {M.}~\bibnamefont
  {Accettulli~Huber}}, \bibinfo {author} {\bibfnamefont {A.}~\bibnamefont
  {Brandhuber}}, \bibinfo {author} {\bibfnamefont {S.}~\bibnamefont
  {De~Angelis}}, \ and\ \bibinfo {author} {\bibfnamefont {G.}~\bibnamefont
  {Travaglini}},\ }\href@noop {} {\  (\bibinfo {year} {2019})},\ \Eprint
  {http://arxiv.org/abs/1911.10108} {arXiv:1911.10108 [hep-th]} \BibitemShut
  {NoStop}%
\bibitem [{\citenamefont {Cano}\ and\ \citenamefont
  {Ruiperez}(2019)}]{Cano:2019}%
  \BibitemOpen
  \bibfield  {author} {\bibinfo {author} {\bibfnamefont {P.~A.}\ \bibnamefont
  {Cano}}\ and\ \bibinfo {author} {\bibfnamefont {A.}~\bibnamefont
  {Ruiperez}},\ }\href {\doibase 10.1007/JHEP05(2019)189} {\bibfield  {journal}
  {\bibinfo  {journal} {JHEP}\ }\textbf {\bibinfo {volume} {05}},\ \bibinfo
  {pages} {189} (\bibinfo {year} {2019})},\ \Eprint
  {http://arxiv.org/abs/1901.01315} {arXiv:1901.01315 [gr-qc]} \BibitemShut
  {NoStop}%
\bibitem [{\citenamefont {Burger}\ \emph {et~al.}(2018)\citenamefont {Burger},
  \citenamefont {Carballo-Rubio}, \citenamefont {Moynihan}, \citenamefont
  {Murugan},\ and\ \citenamefont {Weltman}}]{Burger:2017yod}%
  \BibitemOpen
  \bibfield  {author} {\bibinfo {author} {\bibfnamefont {D.~J.}\ \bibnamefont
  {Burger}}, \bibinfo {author} {\bibfnamefont {R.}~\bibnamefont
  {Carballo-Rubio}}, \bibinfo {author} {\bibfnamefont {N.}~\bibnamefont
  {Moynihan}}, \bibinfo {author} {\bibfnamefont {J.}~\bibnamefont {Murugan}}, \
  and\ \bibinfo {author} {\bibfnamefont {A.}~\bibnamefont {Weltman}},\ }\href
  {\doibase 10.1007/s10714-018-2475-0} {\bibfield  {journal} {\bibinfo
  {journal} {Gen. Rel. Grav.}\ }\textbf {\bibinfo {volume} {50}},\ \bibinfo
  {pages} {156} (\bibinfo {year} {2018})},\ \Eprint
  {http://arxiv.org/abs/1704.05067} {arXiv:1704.05067 [astro-ph.HE]}
  \BibitemShut {NoStop}%
\bibitem [{\citenamefont {Cano}\ and\ \citenamefont
  {Pereñiguez}(2019)}]{Cano:2019ozf}%
  \BibitemOpen
  \bibfield  {author} {\bibinfo {author} {\bibfnamefont {P.~A.}\ \bibnamefont
  {Cano}}\ and\ \bibinfo {author} {\bibfnamefont {D.}~\bibnamefont
  {Pereñiguez}},\ }\href@noop {} {\  (\bibinfo {year} {2019})},\ \Eprint
  {http://arxiv.org/abs/1910.10721} {arXiv:1910.10721 [hep-th]} \BibitemShut
  {NoStop}%
\end{thebibliography}%
\begin{appendix}
\setcounter{equation}{0}
\renewcommand{\theequation}{A.\arabic{equation}}
	\onecolumngrid
	\section{Appendix A: Integral Transforms}
	Here we quote the relevant integral transforms that we make use of in this paper:
	\begin{align}\label{FTs}
	\int \frac{\sd^3\mathbf{q}}{(2\pi)^3}e^{i\mathbf{q}\cdot \mathbf{r}}\frac{1}{\textbf{q}^2} &= \frac{1}{4\pi r}\\
	\int \frac{\sd^3\mathbf{q}}{(2\pi)^3}e^{i\mathbf{q}\cdot \mathbf{r}}\frac{q_i}{\textbf{q}^2} &= \frac{ix_i}{4\pi r^3}\\\
	\int \frac{\sd^3\mathbf{q}}{(2\pi)^3}e^{i\mathbf{q}\cdot \mathbf{r}}|\mathbf{q}|^n &=
	\frac{(n+1)!}{2\pi^2r^{3+n}}\sin\left(\frac{3\pi n}{2}\right),\\
	\int \frac{\sd^3\mathbf{q}}{(2\pi)^3}e^{i\mathbf{q}\cdot \mathbf{r}}q_i|\mathbf{q}|^n &=
	i\frac{(n+1)!(3+n)x_i}{2\pi^2r^{5+n}}\sin\left(\frac{3\pi n}{2}\right),\\
	\int \frac{\sd^4q}{(2\pi)^3}\delta(q^0)\delta(\gamma q^1 - \beta\gamma q^3)e^{i\mathbf{q}\cdot \mathbf{b}}q^\mu|\mathbf{q}|^n &= -\frac{i}{2\pi|\beta\gamma|}\cl{H}_1[r^{n+1}]\hat{b}^\mu,\\
	\cl{H}_\nu[r^n] &= \int_0^\infty r^{n+1} J_\nu(kr) = \frac{2^{n+1}}{k^{n+2}}\frac{\Gamma(\frac12(2 + \nu + n))}{\Gamma(\frac12(\nu - n))}.  
	\end{align}
\end{appendix}
\end{document}